\documentclass[12pt,preprint,useAMS,usenatbib,usegraphicx]{aastex}

\shortauthors{CASTRO, GIZIS, & GAGN\'E}
\shorttitle{{\it CHANDRA} OBSERVATION OF 2MASSW J1139511-315921}

\begin{document}

\title{A {\it CHANDRA} OBSERVATION OF THE TW HYDRAE ASSOCIATION BROWN DWARF 2MASSW J1139511-315921}

\author{Philip J. Castro\altaffilmark{1} and John E. Gizis\altaffilmark{1}}
\affil{Department of Physics and Astronomy, University of Delaware, Newark, DE 19716; pcastro@udel.edu, gizis@udel.edu}
\author{Marc Gagn\'e\altaffilmark{2}}
\affil{Department of Geology and Astronomy, West Chester University, West Chester, PA 19383; mgagne@wcupa.edu}

\begin{abstract}
We report on a sequence of {\it Chandra} X-ray Observatory observations of the TW Hydrae brown
dwarf (BD) 2MASSW J1139511-315921 (2M1139).
In the combined 31 ks ACIS-S exposure, 2M1139 is detected at the $3\sigma$ confidence level.
We find an X-ray luminosity of $L_{\rm X}=1.4^{+2.7}_{-1.0}$ x $10^{26}$ ergs s$^{-1}$ 
or $\log L_{\rm X}/L_{\rm bol}=-4.8\pm0.3$. This object is similar to another 
TW Hydrae BD member, CD-33 7795B (TWA 5B): both have H$\alpha$ emission, both 
show no signatures of accretion, and both have comparable ages and spectral types. 
TWA 5B was previously detected in X-rays with a luminosity of $L_{\rm X}=4$ x $10^{27}$ ergs s$^{-1}$ 
or $\log L_{\rm X}/L_{\rm bol}=-3.4$, an order of magnitude more luminous
in X-rays than 2M1139. We find that the discrepancy between the X-ray luminosity of 
2M1139 and TWA 5B is consistent with the spread in X-ray luminosity in the Orion 
Nebula Cluster (ONC) for BDs of similar spectral types. Though rotation may play a 
role in the X-ray activity of ultracool dwarfs like 2M1139 and TWA 5B, the discrepancy 
cannot be explained by rotation alone. We also examine two X-ray bright objects in the 
FOV of our {\it Chandra} observations and find one to be of spectral type K0IV and identify 
it as a possible RS Canum Venaticorum (RS CVn), and another X-ray bright object whose 
light-curve clearly shows the decay phase of an X-ray flare.
\end{abstract}


\section{INTRODUCTION}
Brown dwarfs are objects whose mass is too low to sustain hydrogen fusion, but massive 
enough to fuse deuterium. With $M$ in the range $0.01-0.075 M_{\odot}$, brown dwarfs lay in the mass 
gap between stars and planets. Objects in this mass range are fully convective, like all stars 
below $0.35 M_{\odot}$ \citep{ChabrierBaraffe1997}.
Solar mass stars generate large-scale magnetic fields by amplifying the small-scale fields 
at the boundary layer (the tachocline) of the radiation zone and the convective envelope 
via differential rotation. The $\alpha-\Omega$ dynamo mechanism was first proposed by \citet{Parker1955} 
and its general features have been confirmed by helioseismology \citep{Charbonneau2010}.
Some fully convective objects are known to have strong chromospheric H$\alpha$ emission, 
which is a signature of magnetic activity \citep{Browning2008}.
Since fully convective objects, low mass stars and brown dwarfs, don't have a boundary layer 
to generate large-scale magnetic fields like their more massive counterparts through the 
$\alpha-\Omega$ dynamo, there must be some other mechanism responsible for the generation of 
magnetic fields in these objects. 
A turbulent dynamo which creates small-scale magnetic fields has been suggested 
by \citet{Durney1993}, where other mechanisms that create large-scale magnetic fields
have been suggested such as the $\alpha^2$ and $\alpha^{2}-\Omega$ dynamo \citep{ChabrierKuker2006}.

The first X-ray detection of a brown dwarf was Cha H$\alpha$ 1 by \citet{NeuhauserComeron1998},
who suggested that the quiescent X-ray emission was a result of a magnetically supported corona.
Since then BDs have been studied and detected in X-rays, field BDs such as LP 944-20; \citep{Rutledgeetal2000}, 
Gl 569 Bab; \citep{Stelzer2004}, and BDs in various star forming regions: ONC;
\citep{Feigelsonetal2002,Preibischetal2005b}, Taurus; \citep{Grossoetal2007}, Chamaeleon;
\citep{Neuhauseretal1999}, Pleiades; \citep{BriggsPye2004}, Rho Ophuicus; \citep{Imanishietal2001,Gagneetal2004}, 
IC 348; \citep{PreibischZinnecker2002}, and TWA 5B; \citep{Tsuboietal2003}.
Many mid to late substellar M dwarfs have observed H$\alpha$ emission. H$\alpha$ is 
indicative of chromospheric activity and accretion, with a steep drop at M9 and 
later \citep{MohantyBasri2003,Gizisetal2000}. It has been proposed
that the chromosphere is heated by the overlying corona (indicated by X-ray emission) 
\citep{PreibischZinnecker2002,Flemingetal1988}. This results in a physical relation between the 
chromosphere and the corona \citep{Tsuboietal2003,GizisBharat2004,Cram1982}. 
During this time a puzzling picture has emerged for X-ray emission from BDs, with some BDs being 
detected in X-rays and others not.

The TW Hydrae association has four confirmed BDs: 2MASSW J1207334-393254 (2M1207),
2MASSW J1139511-315921 (2M1139), CD-33 7795B (TWA 5B), and SSSPM J1102-3431 (SSSPM 1102).
The M8 BD 2M1139, and the M8 BD 2M1207, were discovered by \citet{Gizis2002} during
a survey of the TW Hydrae association in search of BDs. TWA 5B, an M8.5 BD,
was discovered as a companion to TWA 5A at a distance of approximately $2^{\prime \prime}$
by \citet{Lowranceetal1999} using the NIMCOS coronagraph on the Hubble Space Telescope (HST).
SSSPM 1102, an M8.5 BD, was discovered by \citet{Scholzetal2005}.
DENIS J124514.1-442907 (DENIS 1245) is the fifth substellar member within the line of
sight of TW Hydrae but has yet to be confirmed by astrometry as a member of the 
association \citep{Looperetal2007}. This association is an ideal place to study BDs 
and their X-ray emission, with its close proximity of $\sim50$ pc \citep{Mamajek2005}, 
an age of $\sim8-10$ Myr \citep{Songetal2003,Webbetal1999}, and a small sample of well 
studied bona fide BDs. These four confirmed BDs are about the same age, and have 
comparable spectral types, all ranging within half a spectral subtype of M8.5. 
They provide a laboratory to study X-ray emission from BDs with the variables of mass, age, and 
spectral type (effective temperature) as a control.

This paper reports on the X-ray observations of the bona fide BD of TW Hydrae, 2M1139, 
providing all of the confirmed BDs in TW Hydrae with either an X-ray detection or an 
upper limit. We will first discuss the observations, source detection, and astrometry (section 2), 
our discussion (section 3), and finally our conclusions (section 4). In the appendix we provide a 
brief discussion of two X-ray bright objects in the FOV of our {\it Chandra} observations.

\section{OBSERVATIONS AND CIAO DATA REDUCTION}
\subsection{{\it Chandra} Observations}
{\it Chandra} observed 2M1139 in a sequence of three observations in cycle 9, sequence number 200494. 
The first observation, \dataset [ADS/Sa.CXO#Obs/9835] {{\it Chandra} observation ID (ObsId) 9835},
was observed on 2008 March 13th at 14:32 UT until 17:14 UT, having a live-time on CCD 7 of 8.0 ks (2.2 hrs).
The second observation, \dataset [ADS/Sa.CXO#Obs/8913] {{\it Chandra} ObsId 8913},
was observed on 2008 March 17th at 6:01 UT until 8:19 UT, having a live-time on CCD 7 of 7.1 ks (2.0 hrs).
The third observation, \dataset [ADS/Sa.CXO#Obs/9841] {{\it Chandra} ObsId 9841},
was observed on 2008 August 19th at 1:40 UT until 6:35 UT, having a live-time on CCD 7 of 15.8 ks (4.4 hrs).
The roll angle of ObsId 9841 differs from the common roll angle of the other ObsIds by $-145^{\circ}$,
and the aim point is offset from the common aim point of the other ObsIds by $\sim0.5^{\prime}$.
All three observations were obtained in `faint' format, and `timed exposure' mode, with a time resolution 
of 3.2 s using the Advanced CCD Imaging Spectrometer (ACIS) with chips S1-S4, and I2-I3 enabled. 
The target falls on chip S3 so our discussion will only refer to 
this chip. Chip S3 is a back-illuminated CCD with dimensions 1024 x 1024 pixels, spanning a FOV of 
$8^{\prime}$ x $8^{\prime}$, with a pixel size of 24.0 $\mu$m with a spatial resolution at aim point 
of $0\farcs492\pm0\farcs0001$. The back-illuminated S3 chip is better for detecting low energy sources 
than its front-illuminated counterpart \citep{Weisskopfetal2002}, I0-I3, this is due to its larger effective 
area at lower energies (higher sensitivity). Further details regarding {\it Chandra} can be found 
in the {\it Chandra} Proposers' Observatory Guide\footnote{Available at http://cxc.harvard.edu.}.

\subsection{CIAO Data Reduction}
We applied the standard data reduction pipeline using CIAO software version 4.0.2.
Acis\_process\_events was run on each ObsId evt1 fits file. 1) Pixel randomization was removed, 2) Filtered
for bad grades and a clean status column, grades=0,2,3,4,6 with a status=0 were kept, 3)
Good Time Interval's were applied (GTI) which resulted in an evt2 fits file. Finally, the 
energy was filtered from 0.1-8 keV keeping in line with \citet{Tsuboietal2003}. 

\subsection{Image Creation and Source Detection}
All three observations of sequence 200494, ObsId 9835, ObsId 8913, and ObsId 9841, 
were merged using the CIAO routine {\sc merge\_all}. An image of the merged observations was 
created with dimensions 1950 x 1600 pixels. {\sc Wavdetect} was run on the image with 
scales 1, 2, 4, 8, and 16, and with a significance threshold of 5 x $10^{-7}$. Given that the merged 
observations have an area of approximately 2 x $10^{6}$ pixels$^{2}$ this should yield one 
false detection over the entire image. {\sc Wavdetect} yielded 25 source detections over the 
entire image, a cutoff of 1 net counts was made and one source was eliminated. 2M1139 was not 
detected in the full band image.

ACIS counts in the energy bands (0.1-1.5 keV), (1.5-2.5 keV), and (2.5-8 keV) were used to 
create RGB images. {\sc Wavdetect} was run on these images with the same scales and significance
threshold as before. {\sc Wavdetect} detects a source in the soft band image, 0.1-1.5 keV, at (J2000)
R.A.$=11^{h}39^{m}51^{s}.081$, Decl.$=-31^{\circ}59^{\prime}21\farcs75$, $\sim0\farcs25$ from 
the position of 2M1139 (as discussed in section 2.5), albeit with a 1D region due to the small 
number of counts combined with the spatial distribution of the photons. 2M1139 was not detected 
in the medium or hard images. The 3 counts from 2M1139 are from ObsId 9841 only.
In order to assess the reliability of the {\sc wavdetect} 2M1139 soft band detection, Acis 
Extract (AE) was applied to the observations.

\subsection{ACIS Extract}
ACIS Extract (AE), a software package from Penn State \citep{Broosetal2010}, was used to perform 
an optimal source and background extraction, while accounting for the {\it Chandra} point 
spread function (PSF). ACIS Extract was run on the 3 ObsIds after the CIAO data reduction was 
completed. AE only used ObsId 9841 for 2M1139 due to the 3 photons being present only in this ObsId. 
AE found a detection with net counts of $2.84^{+2.94}_{-1.63}$ in the full band (0.5-8 keV), and
net counts of $2.95^{+2.94}_{-1.63}$ in the soft band (0.5-2 keV), where the errors are $1\sigma$ 
upper and lower values. AE determined a `prob\_no\_source' for the sources in the observations, 
`prob\_no\_source' is a statistical measurement of the probability that a source could be a Poisson 
fluctuation in the background, and provides a classic confidence level. In the full band, a `prob\_no\_source' 
of 6.3 x $10^{-4}$ was found for the null hypothesis, this yields a detection at the $3\sigma$ confidence 
level. For the soft band, a `prob\_no\_source' of 1.9 x $10^{-5}$ was found for the null hypothesis, 
this yields a detection at the $4\sigma$ confidence level. The statistics in the soft band are more 
robust since all 3 photons are in this energy range and the background is reduced considerably in 
this energy band. With 3 photons it is not possible to determine whether this detection of 
X-ray emission is variable. Considering that the two shorter observations with no photons for 2M1139, 
ObsId 9835 and ObsId 8913, have exposure times that are approximately the spacing in photon arrival 
times from ObsId 9841, $\sim5$ ks and $\sim7$ ks, variability shouldn't be assumed for ObsId 9841.
\textit{We conclude that we have a $3\sigma$ detection of X-ray emission at the location of 2M1139.} 

\subsection{Astrometry}
We performed astrometry on the {\it Chandra} observations using an I-band image whose coordinate system 
was calibrated using the 2MASS catalog. The 2MASS catalog contained too few counterparts to {\it Chandra} 
sources to confidently perform astrometry directly using the 2MASS catalog.

2M1139 was observed in the I-band on March 24, 2007 with an exposure time of 1200 seconds with 
the 1m telescope at the CTIO. Basic data reduction techniques were applied using IRAF.
Astrometry was performed on the I-band image using the 2MASS catalog by applying IRAF astrometry 
routines, `daofind', `ccxymatch', and `ccmap'. This resulted in a table of sources in the FOV of {\it Chandra} 
larger than that of 2MASS due to the I-band observation being deeper than the 2MASS observations.

We performed astrometry on the {\it Chandra} observations using the data set of I-band sources whose 
positions were calibrated using the 2MASS catalog. We match the data set of I band sources to the {\it Chandra} 
sources found by {\sc wavdetect} in the full band (0.5-8 keV), in the same manner as discussed above.
We find 5 matches with a conservative matching tolerance of $0\farcs5$, with an RMS of $0\farcs036$ 
in the R.A. and $0\farcs115$ in the Decl. After astrometry was performed on the {\it Chandra} observations, 
we find the {\it Chandra} X-ray detection at the location of 2M1139 is $\sim0\farcs25$ from the 
2M1139 source in the I-band image. We use the I-band image as the best estimate of position for 2M1139 
due to the proximity in time of the {\it Chandra} observations (2008) and the I-band observation (2007). 
This was chosen rather than the alternative of using the 2MASS position of 2M1139 and extrapolating the position 
for the epoch of the {\it Chandra} observations (2008) from the proper motion found by \citet{Scholzetal2005}.
\textit{Hence, we claim that the detection of 3 photons at the location of 2M1139 is coincident with 
2M1139 and thus is a bona fide detection of X-ray emission from 2M1139}. 

2M1139 was found to have an absorbed X-ray flux of 5.5 x $10^{-16}$ ergs cm$^{-2}$ s$^{-1}$ 
from the AE output parameter `flux2'. Using the distance of $R=46$ pc determined photometrically
by \citet{Teixeiraetal2008}, and $L_{\rm X}=4\pi R^{2}F$ where $F$ is the flux and $R$ is the 
distance to the target, we determine the X-ray luminosity of 2M1139 to be
$L_{\rm X}=1.4^{+2.7}_{-1.0}$ x $10^{26}$ ergs s$^{-1}$ or $\log L_{\rm X}/L_{\rm bol}=-4.8\pm0.3$.
In determining the error of $L_{\rm X}$ we have assumed an error in distance of 20\% and 
used the $1\sigma$ error in net counts from AE. In determining the error in 
$\log L_{\rm X}/L_{\rm bol}$ we used the $1\sigma$ error in net counts from AE only, 
since $\log L_{\rm X}/L_{\rm bol}$ is independent of distance. 2M1139 has a median energy of 1 keV,
where TWA 5B has a peak energy of 0.7 keV.

2M1139 was included unsuccessfully in a search for planetary mass companions along with other TW 
Hydrae objects, HST proposal ID 10176. It is then unlikely that this {\it Chandra} source may 
be attributed to a companion or background object to within the resolving power of the NICMOS 
instrument, the NIC2 has a scale of $0\farcs075$ pixel$^{-1}$. 

Although 2M1139 is near the threshold of detection of {\it Chandra}, its identification as 
a detection rather than an upper limit is consistent with the source detection criteria of 
other publications. \citet{Townsleyetal2006, Broosetal2007} use the criteria `prob\_no\_source' $<0.003$ 
in the full band (0.5-8 keV) as a detection threshold in determining primary sources. 
\citet{Wangetal2007, Wangetal2010} uses a more conservative criteria of `prob\_no\_source' $<0.001$ 
in the full band (0.5-8 keV) as a detection threshold in determining primary sources. The 
`prob\_no\_source' for 2M1139 in the full band, 6.3 x $10^{-4}$, is well within the above 
criteria. \citet{Townsleyetal2006} include several sources in their primary source list that 
like 2M1139 have slightly less than 3 net counts in the full band. A similar detection of 
a low count source was made by \citet{Audardetal2007}, Kelu-1 an L brown dwarf binary was 
considered a detection with 4 counts in a 24 ks observation by {\it Chandra}, a count 
rate smaller than that of 2M1139. Regardless of the interpretation of the data, the 
question still remains as to the discrepancy of the X-ray luminosities between 2M1139 and 
TWA 5B that are both active as indicated by their H$\alpha$ emission.

\section{DISCUSSION}

2M1207 and SSSPM 1102 both have strong H$\alpha$ emission, as shown in Figure 1, since they both 
have H$\alpha$ emission which may be either from a chromosphere or from accretion, we expect they may 
also have X-ray emission. \citet{Stelzeretal2006} suggested that accretion may suppress X-ray 
emission in BDs, where \citet{Drakeetal2009} suggested the opposite, that coronal X-rays 
modulate the accretion flow in T Tauri stars. Both these mechanisms are consistent with the 
findings by \citet{Preibischetal2005a} for T Tauri stars in the ONC in which they found an 
anti-correlation between mass accretion rate and X-ray activity.

\begin{deluxetable}{lccccccc}
\tabletypesize{\tiny}
\tablecaption{Brown Dwarfs in TW Hydrae}
\tablewidth{0pt}
\tablehead{
\colhead{Name} & \colhead{ST} & \colhead{$v \sin i$} & \colhead{H$\alpha$ EW} & \colhead{H$\alpha$ 10\% FW} & \colhead{$L_{\rm X}$} & \colhead{$\log(L_{\rm X}/L_{\rm BOL})$} & \colhead{References} \\
\colhead{} & \colhead{} & \colhead{(km s$^{-1}$)} & \colhead{(\AA)} & \colhead{(km s$^{-1}$)} & \colhead{(10$^{27}$ ergs s$^{-1}$)} & \colhead{} & \colhead{}
}
\startdata
2M1139 & M8\phantom{.5} & 25$\pm$2 & 9.7, 7.3, 22$\pm$2 & 111$\pm$10 & \phantom{$<$}0.14$^{+0.27}_{-0.10}$ & \phantom{$<$}-4.8$\pm0.3$ & 1, 2, [1,2,3], 2, 4, 4\\
2M1207 & M8\phantom{.5} & 13$\pm$2 & 27.7, 49$\pm$2, 126 &  170-320 & $<$0.12\phantom{$^{+0.00}_{-0.00}$} & $<$-4.8\phantom{$\pm0.0$} & 1, 2, [2, 3, 5], 6, 7, 7\\
TWA 5B & M8.5 & 16$\pm$2 & 20, 5.1 & 162$\pm$10 & \phantom{$<$}4\phantom{.00$^{+0.00}_{-0.00}$} & \phantom{$<$}-3.4\phantom{$\pm0.0$} & 8, 2, [9, 2], 2, 8, 8\\
SSSPM 1102 & M8.5 & $\cdots$ & 10, 64$\pm$3, 50 & 194 & $<$0.87\phantom{$^{+0.00}_{-0.00}$} & $<$-4.0\phantom{$\pm0.0$} & 10, $\cdots$, [10, 3, 5], 11, 6, 4\\
\enddata
\tablerefs{
(1) \citet{Gizis2002}; (2) \citet{Mohantyetal2003}; (3) \citet{Looperetal2007}; (4) this work; (5) \citet{Herczegetal2009}; (6) \citet{Stelzeretal2007}; (7) \citet{GizisBharat2004}; (8) \citet{Tsuboietal2003}; (9) \citet{Neuhauseretal2000}; (10) \citet{Scholzetal2005}; (11) \citet{Scholzetal2005b}.}
\tablecomments{The multiple measurements of H$\alpha$ EW are listed in chronological order of the cited work. We determined the upper limit on $L_{\rm X}$ of SSSPM 1102 by converting the upper limit on $L_{\rm X}$ given by \citet{Stelzeretal2007} which was based on $d=43$ pc to an upper limit based on $d=55.2$ pc determined from trigonometric parralax by \citet{Teixeiraetal2008}. We used BC$_{J}$ from \citet{Wilkingetal1999}, $m_{J}$ from \citet{Scholzetal2005}, and the distance from \citet{Teixeiraetal2008} in determining $L_{\rm bol}$ for 2M1139 and SSSPM 1102.}
\end{deluxetable}

\begin{figure*}
\caption{H$\alpha$ EW measurements for the four BDs of TW Hydrae. The stars represent the
X-ray detections of 2M1139 and TWA 5B \protect\citep{Tsuboietal2003}, and the arrows
represent upper limits. Multiple measurements are listed according to those found in the
literature, refer to Table 1.
}
\includegraphics[width=6.in]{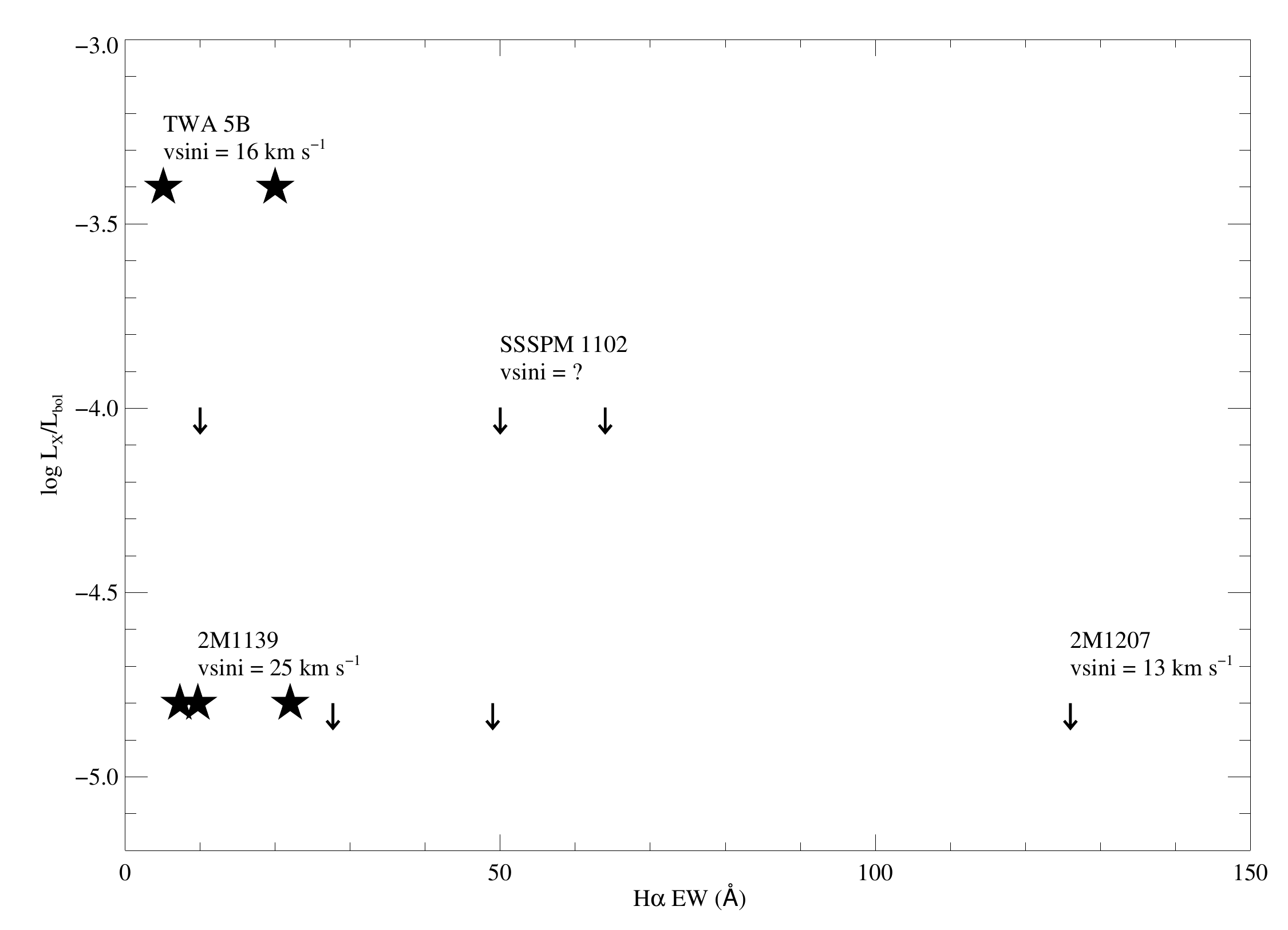}
\end{figure*}

\begin{figure*}
\caption{H$\alpha$ 10\% FW for the four BDs of TW Hydrae. The symbols are the same as in Figure 1.
The values of H$\alpha$ 10\% FW spanned by 2M1207 show the range due to variability found
by \protect\citet{Stelzeretal2007}. The dashed vertical line separates the values of H$\alpha$ 10\% FW
for an object to be considered an accretor (non-accretor) according to \protect\citet{Jayawardhanaetal2003}.
}
\includegraphics[width=6.in]{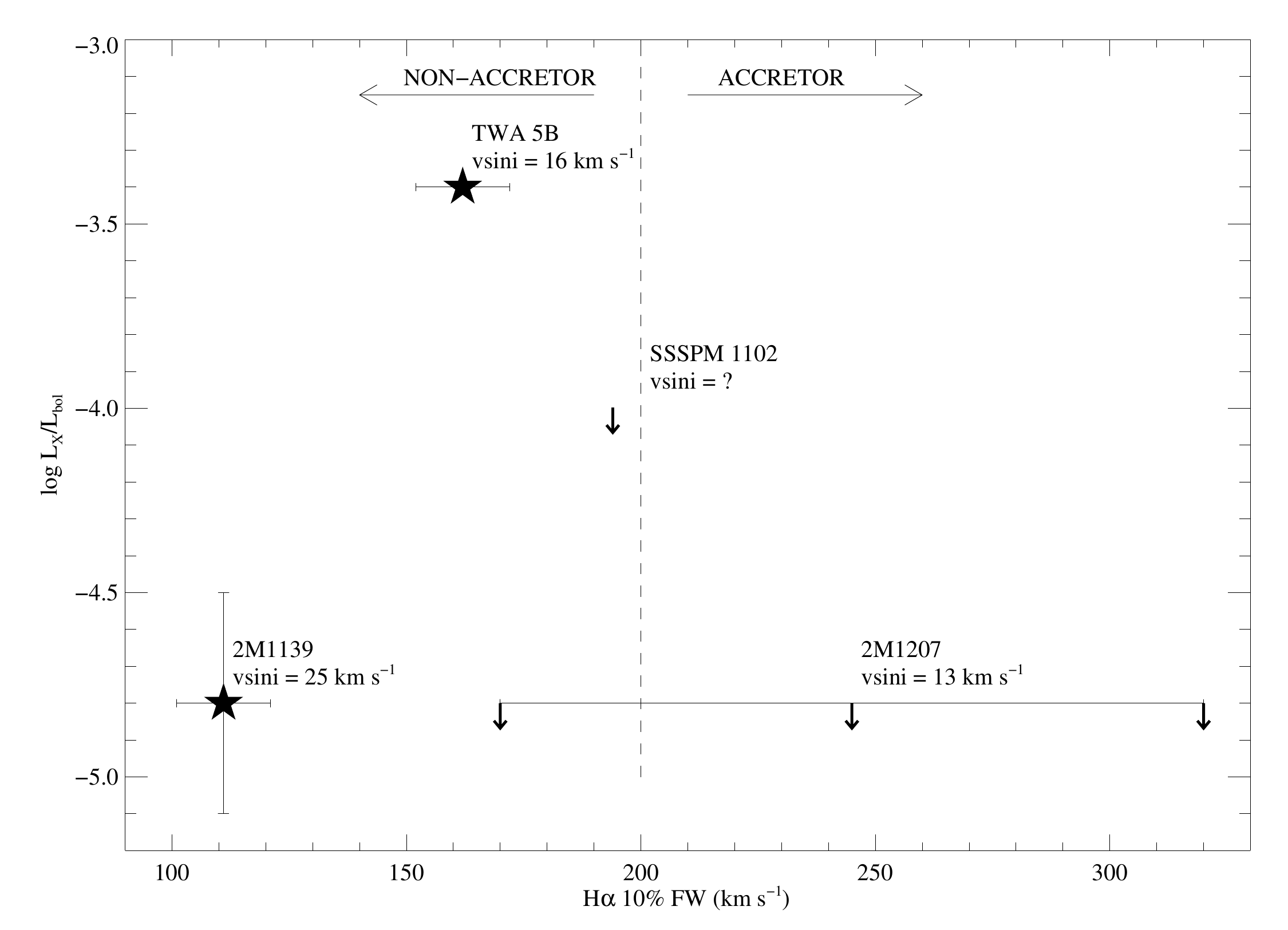}
\end{figure*}

H$\alpha$ EW is a common criteria in distinguishing 
accreting, classical T Tauri stars (CTTS), from non-accreting, weak-lined T Tauri stars (WTTS).
However, H$\alpha$ EW has no unique cutoff that distinguishes all CTTS from WTTS. This is due to 
the `contrast effect', different spectral types have different continuums for the same level of 
chromospheric saturation, causing H$\alpha$ lines to appear more prominent in one spectra type 
than another \citep{WhiteBasri2003}. \citet{WhiteBasri2003} proposed a different diagnostic 
criterion in distinguishing CTTS (accreting) from WTTS (non-accreting). They propose using the 
H$\alpha$ 10\% FW and a cutoff of 270 km s$^{-1}$, an H$\alpha$ 10\% FW $>270$ km s$^{-1}$ is 
considered a CTTS (accretor). This criterion, 10\% of full width, is independent of the 
continuum level of stars and thus of the spectral type, and may be a more accurate diagnostic of 
accretion than H$\alpha$ EW \citep{WhiteBasri2003}. \citet{Jayawardhanaetal2003} has suggested a 
modification to the criteria of \citet{WhiteBasri2003}, they adopted an H$\alpha$ 10\% FW of 
$\gtrsim200$ km s$^{-1}$ as a criteria for accretion for low mass stars and brown dwarfs. 
They adopt this criterion over that of \citet{WhiteBasri2003} because the criteria of $>270$ km s$^{-1}$ 
can fail for very low mass objects with low accretion rates \citep{Jayawardhanaetal2003}. We 
thus adopt the criteria of $\sim200$ km s$^{-1}$ as a cutoff for accretion, substellar objects 
with $\gtrsim200$ km s$^{-1}$ are considered accretors.

2M1207 is considered an accreting substellar object due its large H$\alpha$ 10\% FW emission 
values (as shown in Figure 2), where SSSPM 1102 is considered an accreting substellar object 
due its large H$\alpha$ 10\% FW emission value along with measured accretion rates.
2M1207 was found to have variable H$\alpha$ 10\% FW emission by \citet{Stelzeretal2007}, 
with values ranging from 170-320 km s$^{-1}$, where SSSPM 1102 has an H$\alpha$ 10\% FW 
of 194 km s$^{-1}$, putting it on the border of accreting. 2M1207 was found to have an 
accretion rate of 1.3 x $10^{-12}$ $M_{\odot}$/yr, and SSSPM 1102 was found to have an 
accretion rate of 1.6 x $10^{-13}$ $M_{\odot}$/yr, both from Balmer emission by 
\citet{Herczegetal2009}. 2M1207 was observed by {\it Chandra} in a 50 ks observation 
by \citet{GizisBharat2004} with no detection, being given an upper limit of 
$L_{\rm X}<1.2$ x $10^{26}$ ergs s$^{-1}$ or $\log L_{\rm X}/L_{\rm bol}<-4.8$. 
\citet{GizisBharat2004} concluded that the H$\alpha$ emission was from star-disk 
interactions (accretion). SSSPM 1102 was found in an archival XMM Newton observation 
by \citet{Stelzeretal2007}, it was not detected and was given an upper limit of 
$L_{\rm X}<8.7$ x $10^{26}$ ergs s$^{-1}$ or $\log L_{\rm X}/L_{\rm bol}<-4.0$. 
These X-ray non-detections are consistent with the idea that mass accretion rate and X-ray 
activity are anti-correlated.

TWA 5B was found to have moderate H$\alpha$ emission as shown in Figure 1 and Table 1.
TWA 5B is not considered an accretor, with an H$\alpha$ 10\% FW of 162 km s$^{-1}$.
TWA 5B was observed and detected in X-rays in a 10.3 ks observation with {\it Chandra} by 
\citet{Tsuboietal2003}. \citet{Tsuboietal2003} detected 35 photons, and determined the 
X-ray emission to be quiescent with an X-ray luminosity of $L_{\rm X}=4$ x $10^{27}$ ergs s$^{-1}$ 
or $\log L_{\rm X}/L_{\rm bol}=-3.4$.

2M1139 also has moderate H$\alpha$ emission as shown in Figure 1 and Table 1, values that 
are quite similar to TWA 5B. It is not an accretor with an H$\alpha$ 10\% FW of 111 km s$^{-1}$, 
the lowest of the four BDs in TW Hydrae. We detect X-ray emission from 2M1139 as discussed in section 2.5.

It is not surprising that 2M1139 and TWA 5B have similar H$\alpha$ EW measurements since 
chromospheric H$\alpha$ activity saturates at $v \sin i \approx10$ km s$^{-1}$ in 
M5.5-M8.5 dwarfs \citep{MohantyBasri2003}, and both 2M1139 and TWA 5B have $v \sin i$ values 
greater than this. Where the H$\alpha$ EW measurements of 2M1207 and SSSPM 1102 are greater 
than and equal to those of 2M1139 and TWA 5B due to accretion and (or) chromospheric activity.

The question is then why does TWA 5B have an X-ray luminosity $\sim10$ times greater than 
2M1139. \citet{Preibischetal2005b} found in the ONC there is an intrinsic range of $\log L_{\rm X}/L_{\rm bol}$ 
for BDs of similar spectral type, with values from $\log L_{\rm X}/L_{\rm bol}$ $\approx$ $-3.3$ to $-4.2$
for BDs that are of spectral type M6.5-M9, a spread about an order of magnitude. \citet{Bergeretal2010} 
found from an up to date sample of ultra cool dwarfs, a dispersion of $\log L_{\rm X}/L_{\rm bol} 
\approx -5$ to $-3$ for M7-M9, where 2M1139 and TWA 5B are well within this range. At this point 
we may conclude that there is an intrinsic range of X-ray luminosity from BDs of similar spectral 
types, and 2M1139 and TWA 5B span the lower and upper values of this range. A characteristic that 
is largely responsible for the range of $\log L_{\rm X}/L_{\rm bol}$ for quiescent X-ray emission 
in main sequence stars with an $\alpha-\Omega$ dynamo is the equatorial rotation velocity, more 
accurately the Rossby number which is a function of the equatorial rotation velocity \citep{Feigelsonetal2003}.
The characteristics responsible for this intrinsic range in fully convective stars is still not understood.

We can rule out flaring as an explanation for the large X-ray emission of TWA 5B compared to 2M1139.
Flaring is unlikely considering that \citet{Tsuboietal2003} found it to be quiescent over a 10 ks 
observation. TWA 5B does not show the rise and decay of a flare. \citet{Bergeretal2010} found 
ultracool dwarfs to have typical flaring durations of approximately 1 hour, with a range of about 
$10^{2}$ to $10^{4}$ seconds, thus the rise or decay of a flare from TWA 5B would probably have been 
seen during the 10 ks observation. 

We compare and contrast 2M1139 and TWA 5B: both have H$\alpha$ emission, both show no signatures of 
accretion, and have comparable ages and spectral types. In an effort to exploit discrepancies between 
the two to understand why one has high X-ray emission near the saturation limit and the other has very 
low X-ray emission, we examine the equatorial rotation velocity and X-ray emission of 2M1139 and TWA 5B.

\subsection{Rotation}
\citet{MohantyBasri2003} looked at fully convective objects,
M4-L6, analyzing the activity-rotation relation for chromospheric activity (H$\alpha$ emission). 
They found that for M4-M8.5 a saturation type activity-rotation relation exists for chromospheric activity 
(H$\alpha$ emission), extending the work on H$\alpha$ activity from \citet{Delfosseetal1998} into the 
late M spectral type. They also find a drastic drop in H$\alpha$ activity for spectral type $\geq$M9. 
This may be the result of very high resistivities in the neutral atmospheres, perhaps combined with the 
rapid formation of dust, which would damp magnetic energy available for supporting a chromosphere 
\citep{MohantyBasri2003}. \citet{MohantyBasri2003} use $v \sin i$ as a measure of rotation, and is 
limited because it only gives a lower limit on the equatorial rotation velocity. Nonetheless, from 
these works it is clear there is a saturation plateau for fully convective objects.

Work by \citet{Bergeretal2008a}, \citet{Bergeretal2008b}, and \citet{Bergeretal2010} are beginning 
to piece together the picture of activity in fully convective objects of spectral type $>$M7. Their 
work finds a distinct change in activity around spectral type $\sim$M7. They find a decline in X-ray 
activity for spectral type $\geq$M7, and a decline in H$\alpha$ activity at $\geq$M6, slightly earlier 
in spectral type than \citet{MohantyBasri2003}.

\citet{Bergeretal2010}, Figure 9, examine the activity-rotation relation of ultracool dwarfs and find 
for moderate rotators, $1>P>0.3$ days, a median $\log L_{\rm X}/L_{\rm bol}=-4$. Fast rotators, 
P$<0.3$ days, are found to generally have weaker X-ray emission, a median $\log L_{\rm X}/L_{\rm bol}=-5$, 
and they suggest that this may be analogous to the super saturation regime for main sequence objects. 
Ultracool dwarfs that are moderate rotators saturate at $\log L_{\rm X}/L_{\rm bol}=-3$, like 
main sequence stars. But unlike the main sequence activity-rotation relation for X-ray emission, 
based on detections there exists a large dispersion below the saturation plateau, from 
$\log L_{\rm X}/L_{\rm bol} \approx -5$ to $-3$. All of the detections are from spectral types M7-M9, 
with the exception of the L2, Kelu-1, with $\log L_{\rm X}/L_{\rm bol}=-4.3$. \citet{Bergeretal2010} 
notes that this is based on a small number of objects and needs to be confirmed with larger numbers, 
but it suggests the possibility of a super saturation regime. A number of possible explanations for 
the super saturation region have been proposed, intrinsic dynamo effects, secondary effects of 
centrifugal stripping, or that the source of coronal heating is different (inefficient heating) 
\citep{Bergeretal2008b}.

\begin{figure*}
\caption{Figure 9 from \protect\citet{Bergeretal2010}, with the data from 2M1139 and TWA 5B of this 
work overlaid as diamond symbols. The four tick marks for 2M1139 and TWA 5B represent, from left 
to right, an inclination of $15^{\circ}$, $30^{\circ}$, $45^{\circ}$, and $90^{\circ}$. The dashed 
line represents the approximate upper envelope of quiescent X-ray detections for this plot, based 
on a small sample size of $\geq$M7 dwarfs. The lower envelope is not well defined based on only 2 
detections and many upper limits. This figure shows that rotation may play a role in the X-ray 
activity of ultracool dwarfs like 2M1139 and TWA 5B, but due to the dispersion of $\log L_{\rm X}/L_{\rm bol}$ 
below the upper envelope, rotation alone cannot explain the difference in X-ray emission between 
2M1139 and TWA 5B.
}
\includegraphics[width=6.in]{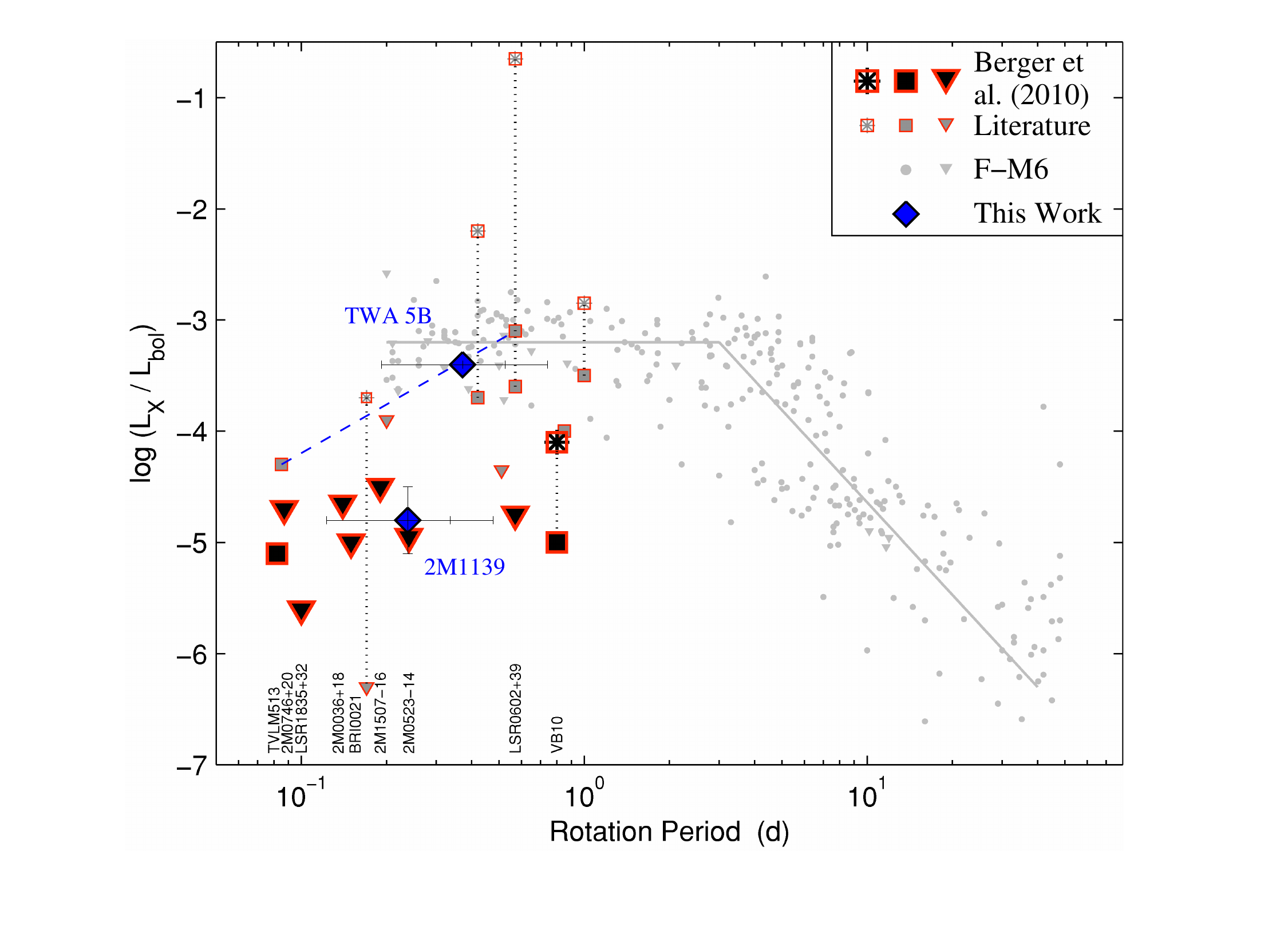}
\end{figure*}

Periods of 2M1139 and TWA 5B are not known, but we can determine a spread of values for period 
given their $v \sin i$ values and a few select inclinations. We use $P=(2\pi R)/$ $v$ where 
$v$ $=V/$sin $i$, $P$ is the period, $R$ is the radius, $v$ is the equatorial rotation velocity, 
$i$ is the inclination, and $V$ is the measured $v \sin i$ value. We used the radius of 2M1207 as 
an estimate for the radii of 2M1139 and TWA 5B in determining the period due to similarities in age 
and mass. We inferred the radius of 2M1207 directly from the $L_{\rm bol}$ and $T_{\rm eff}$ 
from \citet{Mohantyetal2007}.

Figure 3 shows the activity-rotation relation for X-ray activity in ultracool dwarfs, taken from Figure 9 
of \citet{Bergeretal2010}, with the results of this work overlaid as diamond symbols. The four tick 
marks for 2M1139 and TWA 5B show, from left to right, inclinations of $15^{\circ}$, $30^{\circ}$, 
$45^{\circ}$, and $90^{\circ}$. The dashed line shows the approximate upper envelope for quiescent 
X-ray detections for this figure, where the lower envelope is not well defined based on only 2 
detections and many upper limits. X-ray activity is shown to have considerable scatter, 
$-5<\log L_{\rm X}/L_{\rm bol}<-3$ for moderate rotators, P$>0.3$ days, and for the scatter to 
decrease as the period decreases. \textit{The detections of 2M1139 and TWA 5B span this scatter in 
X-ray activity. Though rotation may play a role in the X-ray activity of ultracool dwarfs, the discrepancy 
between 2M1139 and TWA 5B cannot be explained by rotation alone.} 

\citet{Riceetal2010} gives different $v \sin i$ values for 2M1139 and TWA 5B than those determined by 
\citet{Mohantyetal2003}, Table 1. They report a $v \sin i$ of 30 km s$^{-1}$ for 2M1139 and 27 km s$^{-1}$ for 
TWA 5B. The value of $v \sin i$ for TWA 5B is an increase of about 70\% over that of \citet{Mohantyetal2003}. 
Their measurements for the $v \sin i$ values of field objects have disparities with previous measurements 
as well, twice the previous measured value for 2MASS 0140+27 and three times the previous measured value 
for LP 412-31. \citet{Riceetal2010} note that their temperatures measured from spectral fitting for TWA 5B 
and these field objects are hotter than previously determined temperatures. They argue this is likely a 
consequence of increasing importance on FeH absorption for cooler and higher gravity objects that isn't 
reproduced by the atmosphere models they used. Also, that their spectra for TWA 5B has a low signal to noise 
and that it may be contaminated by its bright primary. Considering that \citet{Mohantyetal2003} gives errors 
of $\pm2$ km s$^{-1}$ for their $v \sin i$ measurements, and the issues discussed above for TWA 5B, we thus 
consider the measurements of $v \sin i$ by \citet{Mohantyetal2003} to be more reliable and use these in our 
analysis.

\citet{VaccaSandell2011} find an age for TW Hya, and thus of the coevolved TW Hydrae association, of $\sim3$ Myr.
This age is in contrast to the older accepted age of $\sim8-10$ Myr. There are a number of characteristics 
that may be expected to change with age if these BDs in TW Hydrae are actually $\sim3$ Myr rather than the 
accepted age of $\sim8-10$ Myr. In discussing these characteristics based on age we must exclude TW 
Hydrae BDs from the analysis of cited work since its age is the topic of discussion. According to 
\citet{Osorioetal2006} the projected rotational velocity would be slightly lower because BDs spin 
up with age due to gravitational contraction. From \citet{Stelzeretal2006} the effective temperature 
would be slightly higher. Without the detection of TWA 5B at $\sim10$ Myr, a relation between X-ray
plasma temperature and age becomes ambiguous \citep{Tsuboietal2003}. There appears to be no clear relation 
between age and $\log L_{\rm X}/L_{\rm bol}$ \citep{Stelzer2004,Tsuboietal2003}, except to say that 
we still see moderate X-ray activity at ages of several hundred million years, Gl 569Bab \citep{Stelzer2004}.
No assumptions about the age of TW Hydrae have been made in our analysis of 2M1139 and 
the BDs of TW Hydrae. Furthermore, our analysis comparing and contrasting the BDs of TW Hydrae is age 
independent since they are coevolved. Thus, a younger age of $\sim3$ Myr by \citet{VaccaSandell2011} for 
TW Hydrae has no effect on our conclusions.

\section{CONCLUSIONS}
We report a $3\sigma$ detection of 2M1139 in X-rays with $L_{\rm X}=1.4^{+2.7}_{-1.0}$ x $10^{26}$ ergs s$^{-1}$
or $\log L_{\rm X}/L_{\rm bol}=-4.8\pm0.3$. 2M1139 is similar in H$\alpha$ activity, lack of accretion signatures, 
age, and spectral type to TWA 5B, another BD in TW Hydrae, yet TWA 5B has $\sim10$ times the X-ray luminosity. 
We find the span in $\log L_{\rm X}/L_{\rm bol}$ between these two objects in consistent with BDs of similar 
spectral types in the ONC \citep{Preibischetal2005b}. Based on recent work on the activity-rotation relation 
for X-ray activity of fully convective objects \citep{Bergeretal2010}, we find that rotation may play a role 
in the X-ray activity of ultracool dwarfs like 2M1139 and TWA 5B. However, due to the large span in 
$\log L_{\rm X}/L_{\rm bol}$ of detections, which is comparable to the span of $\log L_{\rm X}/L_{\rm bol}$ 
between 2M1139 and TWA 5B, the discrepancy cannot be explained by rotation alone.

Future work should include X-ray observations of DENIS 1245. If found to be a bona fide member of TW Hydrae, 
further study would help to complete the picture of substellar objects at an age of $\sim10$ Myr. DENIS 1245 
has detected H$\alpha$ emission with an H$\alpha$ EW of 15 \AA$\phantom{0}$\citep{Looperetal2007}, so we 
know it to be active. If DENIS 1245 is found to have X-ray emission, its place in context to the other 
four BDs of TW Hydrae, specifically 2M1139 and TWA 5B, will help to shape our understanding of X-ray activity 
of BDs at the age of $\sim10$ Myr.

\section{ACKNOWLEDGMENTS}
Support for this work was provided by NASA research grant NNG06GJ03G. Support for this work was provided by 
the National Aeronautics and Space Administration through {\it Chandra} Award Number GO89011X issued by 
the {\it Chandra} X-ray Observatory Center, which is operated by the Smithsonian Astrophysical Observatory 
for and on behalf of the National Aeronautics Space Administration under contract NAS8-03060. This work has 
made use of telescopes operated by the SMARTS consortium. This publication makes use of data products from the 
Two Micron All Sky Survey, which is a joint project of the University of Massachusetts and the Infrared Processing 
and Analysis Center/California Institute of Technology, funded by the National Aeronautics and Space Administration 
and the National Science Foundation. This research has made use of the SIMBAD database, operated at CDS, 
Strasbourg, France. This research also makes use of the ACIS Extract software package maintained at Pennsylvania 
State University. We thank Dr. James MacDonald for many useful discussions. We thank the anonymous referee for 
suggestions that helped to improve the manuscript.

\appendix

\section{{\it CHANDRA} X-RAY SOURCES IN THE FOV}
\subsection{Table 2}

\begin{deluxetable}{ccccccccc}
\tabletypesize{\tiny}
\rotate
\tablecaption{{\it Chandra} X-Ray Sources in the FOV}
\tablewidth{0pt}
\tablehead{
\multicolumn{3}{}{} & \multicolumn{3}{c}{Full Band: $0.5-8$ keV} & & \multicolumn{2}{c}{Soft Band: $0.5-2$ keV} \\ \cline{4-6} \cline{8-9}
\colhead{IAU} & \colhead{R.A.} & \colhead{Decl.} & \colhead{Net} & \colhead{Absorbed Flux} & \colhead{Median} & & \colhead{Net} & \colhead{Absorbed Flux} \\
\colhead{Designation} & & & \colhead{Counts} & \colhead{10$^{-16}$} & \colhead{Energy} & & \colhead{Counts} & \colhead{10$^{-16}$} \\
\colhead{(J2000)} & \colhead{(J2000)} & \colhead{(J2000)} & & \colhead{(ergs cm$^{-2}$ s$^{-1}$)} & \colhead{(keV)} & & & \colhead{(ergs cm$^{-2}$ s$^{-1}$)} \\
\colhead{(1)} & \colhead{(2)} & \colhead{(3)} & \colhead{(4)} & \colhead{(5)} & \colhead{(6)} & & \colhead{(7)} & \colhead{(8)}
}
\startdata
J113924.6-315923 & 174.85276 & -31.989982 & $ 75.97^{+ 9.88}_{- 8.81}$ &  340.6 & 1.1 & \phantom{0} & $ 63.25^{+ 9.05}_{- 7.98}$ &  168.6\\
J113926.0-315941 & 174.85865 & -31.994910 & $ 24.05^{+ 6.17}_{- 5.06}$ &  107.1 & 1.4 & \phantom{0} & $ 14.38^{+ 4.97}_{- 3.83}$ &   38.2\\
\textbf{J113926.6-320107} & \textbf{ 174.86101} & \textbf{-32.018644} & $\mathbf{196.08^{+15.10}_{-14.06}}$ & \textbf{ 869.0} & \textbf{1.0} & \textbf{\phantom{0}} & $\mathbf{187.27^{+14.74}_{-13.70}}$ & \textbf{ 495.0}\\
J113931.7-315841 & 174.88215 & -31.978158 & $  6.18^{+ 3.78}_{- 2.58}$ &   26.6 & 3.5 & \phantom{0} & $  0.71^{+ 2.32}_{- 0.83}$ &    1.8\\
J113931.9-315540 & 174.88311 & -31.927813 & $ 15.68^{+ 5.21}_{- 4.08}$ &   69.5 & 1.6 & \phantom{0} & $ 10.63^{+ 4.43}_{- 3.26}$ &   29.9\\
J113942.9-315553 & 174.92897 & -31.931590 & $ 57.85^{+ 8.73}_{- 7.66}$ &  116.7 & 1.0 & \phantom{0} & $ 51.63^{+ 8.26}_{- 7.19}$ &   65.8\\
J113944.9-320249 & 174.93740 & -32.047128 & $ 10.47^{+ 4.43}_{- 3.26}$ &   44.8 & 1.8 & \phantom{0} & $  6.81^{+ 3.78}_{- 2.58}$ &   18.1\\
J113947.2-320010 & 174.94701 & -32.002880 & $  2.84^{+ 2.94}_{- 1.63}$ &    5.6 & 3.7 & \phantom{0} & $  0.95^{+ 2.32}_{- 0.83}$ &    1.1\\
J113947.3-315613 & 174.94744 & -31.936991 & $ 74.39^{+ 9.70}_{- 8.64}$ &  148.1 & 1.4 & \phantom{0} & $ 51.74^{+ 8.26}_{- 7.19}$ &   63.4\\
J113947.6-320024 & 174.94841 & -32.006832 & $ 12.82^{+ 4.71}_{- 3.55}$ &   25.2 & 1.8 & \phantom{0} & $  6.96^{+ 3.78}_{- 2.58}$ &    8.2\\
J113950.6-320038 & 174.96110 & -32.010591 & $ 16.81^{+ 5.21}_{- 4.08}$ &   32.9 & 1.2 & \phantom{0} & $  9.95^{+ 4.28}_{- 3.10}$ &   11.7\\
\textbf{J113951.0-315921} & \textbf{ 174.96284} & \textbf{-31.989374} & $\mathbf{  2.84^{+ 2.94}_{- 1.63}}$ & \textbf{   5.5} & \textbf{1.0} & \textbf{\phantom{0}} & $\mathbf{  2.95^{+ 2.94}_{- 1.63}}$ & \textbf{   3.4}\\
J113951.6-315723 & 174.96539 & -31.956664 & $  4.74^{+ 3.40}_{- 2.15}$ &    9.3 & 3.8 & \phantom{0} & $  0.92^{+ 2.32}_{- 0.83}$ &    1.1\\
J113952.7-315709 & 174.96974 & -31.952770 & $  7.70^{+ 3.96}_{- 2.76}$ &   15.3 & 1.4 & \phantom{0} & $  5.89^{+ 3.60}_{- 2.37}$ &    7.0\\
J113953.9-315830 & 174.97481 & -31.975213 & $ 15.85^{+ 5.09}_{- 3.95}$ &   31.0 & 1.1 & \phantom{0} & $ 12.96^{+ 4.71}_{- 3.55}$ &   15.2\\
J113954.1-315807 & 174.97574 & -31.968710 & $ 40.81^{+ 7.46}_{- 6.37}$ &   79.6 & 1.4 & \phantom{0} & $ 30.94^{+ 6.63}_{- 5.53}$ &   36.3\\
J113956.2-320005 & 174.98438 & -32.001403 & $  6.85^{+ 3.78}_{- 2.58}$ &   13.4 & 1.8 & \phantom{0} & $  3.94^{+ 3.18}_{- 1.91}$ &    4.6\\
J113956.7-315852 & 174.98626 & -31.981228 & $  8.82^{+ 4.12}_{- 2.94}$ &   17.3 & 1.4 & \phantom{0} & $  5.94^{+ 3.60}_{- 2.37}$ &    7.0\\
J113956.8-315623 & 174.98699 & -31.939751 & $ 17.43^{+ 5.33}_{- 4.20}$ &   35.5 & 0.9 & \phantom{0} & $ 15.84^{+ 5.09}_{- 3.95}$ &   19.4\\
J113958.3-320305 & 174.99300 & -32.051401 & $ 11.12^{+ 4.57}_{- 3.41}$ &   23.4 & 1.9 & \phantom{0} & $  5.71^{+ 3.60}_{- 2.37}$ &    7.6\\
J114001.2-315503 & 175.00511 & -31.917502 & $ 28.47^{+ 6.55}_{- 5.45}$ &  111.5 & 1.6 & \phantom{0} & $ 22.68^{+ 5.87}_{- 4.76}$ &   53.3\\
\textbf{J114004.6-320040} & \textbf{ 175.01927} & \textbf{-32.011213} & $\mathbf{780.56^{+28.96}_{-27.94}}$ & \textbf{2950.3} & \textbf{1.0} & \textbf{\phantom{0}} & $\mathbf{734.76^{+28.12}_{-27.10}}$ & \textbf{1700.0}\\
J114014.8-315527 & 175.06200 & -31.924189 & $  4.39^{+ 3.98}_{- 2.79}$ &   18.4 & 4.3 & \phantom{0} & $  1.03^{+ 2.67}_{- 1.30}$ &    2.5\\
J114015.7-315844 & 175.06565 & -31.978890 & $  8.23^{+ 4.28}_{- 3.11}$ &   32.0 & 2.0 & \phantom{0} & $  3.57^{+ 3.18}_{- 1.91}$ &    8.5\\
J114018.3-315654 & 175.07654 & -31.948447 & $ 30.61^{+ 6.90}_{- 5.81}$ &  125.0 & 2.4 & \phantom{0} & $ 10.88^{+ 4.58}_{- 3.42}$ &   26.3\\
\enddata
\tablecomments{The rows in bold highlight 2M1139 and the two objects of interest in the FOV of the {\it Chandra} observations.}
\end{deluxetable}

For completeness we have provided the output from ACIS Extract for the entire FOV of our {\it Chandra} observations. 
Table 2 shows the sources detected by {\sc wavdetect} with scales 1, 2, 4, 8, 
and 16, with a significance threshold of 5 x $10^{-7}$, and 2M1139 which was detected in the soft band image, 0.1-1.5 keV, 
under the same scales and significance threshold. Column 1 is the IAU designation in J2000 for each source, and columns 2 
and 3 are the Right Ascension and Declination in J2000, respectively. Columns 4 to 6 show the 
full band (0.5-8 keV), and 7 and 8 show the soft band (0.5-2 keV) output. Columns 4 and 7 are the net counts,
and columns 5 and 8 are the absorbed flux from the output parameter `flux2'. Column 6 is the median energy, 
`energ\_pct50\_observed', which is the background subtracted median observed energy, where the observed energy 
is based on net counts rather than the `energ\_pct50\_incident' which is based on flux.

\subsection{J114004.6-320040}
An X-ray source in the FOV of our {\it Chandra} observations that warranted further investigation was 
J114004.6-320040, 2MASS source J11400464-3200411. This was the brightest X-ray source in the FOV, having 
almost 800 counts. It was detected by GALEX in the near-ultraviolet (NUV) and far-ultraviolet (FUV).

The light-curve for J114004.6-320040 is shown in Figure 4 (top). The KS statistic yields a probability 
of 0.01, or that there is a 99\% probability that the object is variable. We applied the maximum likelihood 
method to the light-curve of J114004.6-320040 and found that within 95\% confidence that it was variable 
between the first (1.9 hrs) 7 ks and the last (2.6 hrs) 9 ks, due to the count rate jump at (1.9 hrs) 7 ks.

We use the CIAO routine {\sc psextract} and the spectral fitting package XSPEC to fit the spectra of 
J114004.6-320040 (Figure 5, top left) with a two temperature APEC (Wilm abund) model. We find a good fit 
with $kT_{1}=0.73\pm0.04$ keV, $kT_{2}=2.2\pm0.4$ keV, and a column density of 
$N_{\rm H}=1.0\pm0.3$ x $10^{21}$ cm$^{-2}$. We find an unabsorbed flux of 
$f_{\rm xu}=1.5$ x $10^{-13}$ ergs cm$^{-2}$ s$^{-1}$ and an absorbed flux 
of $f_{\rm xa}=1.3\pm0.1$ x $10^{-13}$ ergs cm$^{-2}$ s$^{-1}$.

\begin{figure*}
\caption{Light curves for the two objects of interest in the FOV of the {\it Chandra} observations.
Light-curves are binned over 1 ks. Top: Light-curve for object J114004.6-320040 that was observed
during ObsId 9841 only. Bottom: Light-curve for object J113926.6-320107 from ObsId 9835 (bottom left)
which shows quiescent emission and ObsId 8913 (bottom right) which clearly shows the decay phase of a flare.
}
\includegraphics[width=6.in]{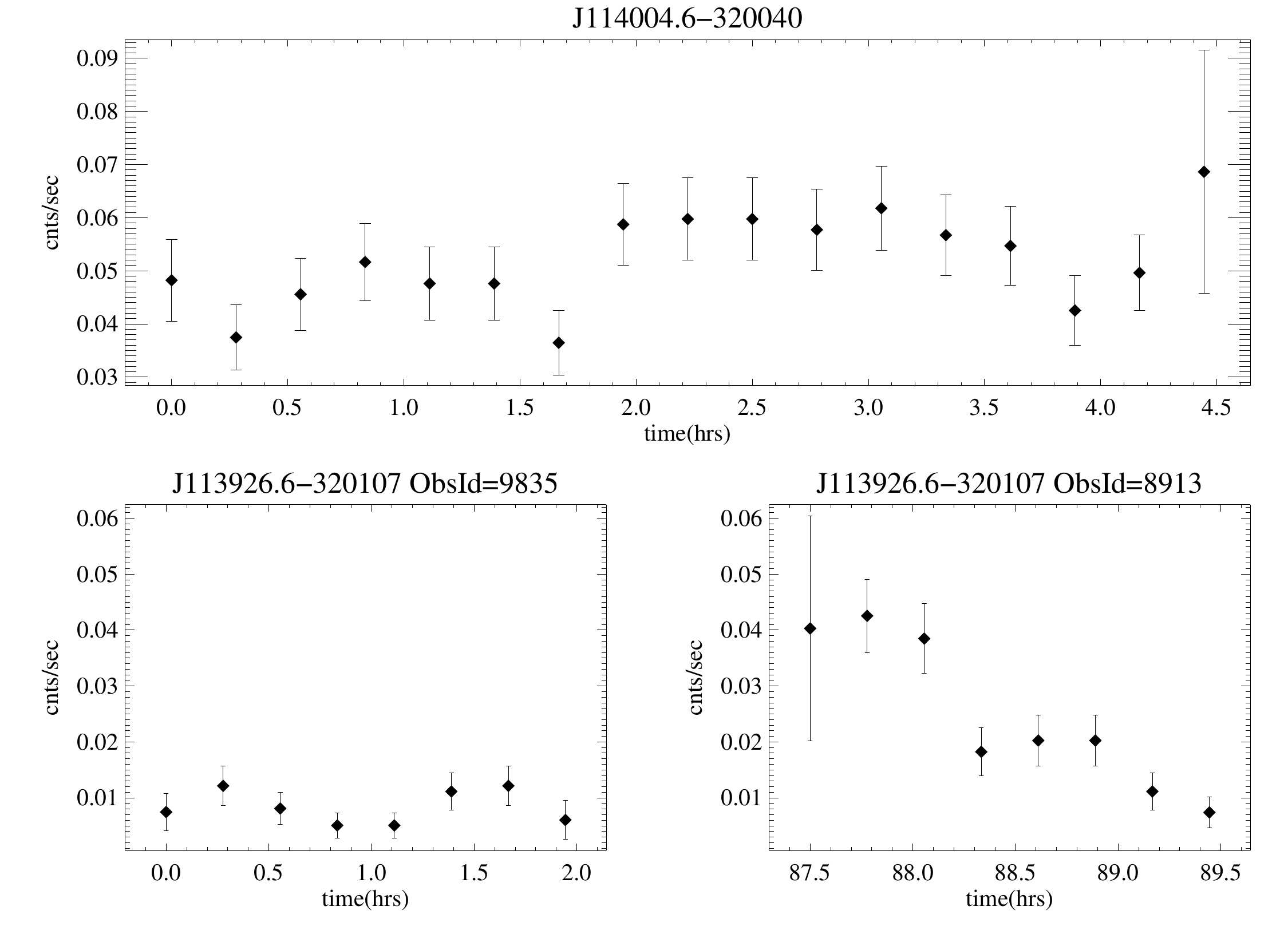}
\end{figure*}

\begin{figure*}
\caption{{\it Chandra} spectra of J114004.6-320040 and J113926.6-320107 for ObsId 9835 (quiescent) and
ObsId 8913 (flaring).
}
\includegraphics[width=2.5in,angle=270]{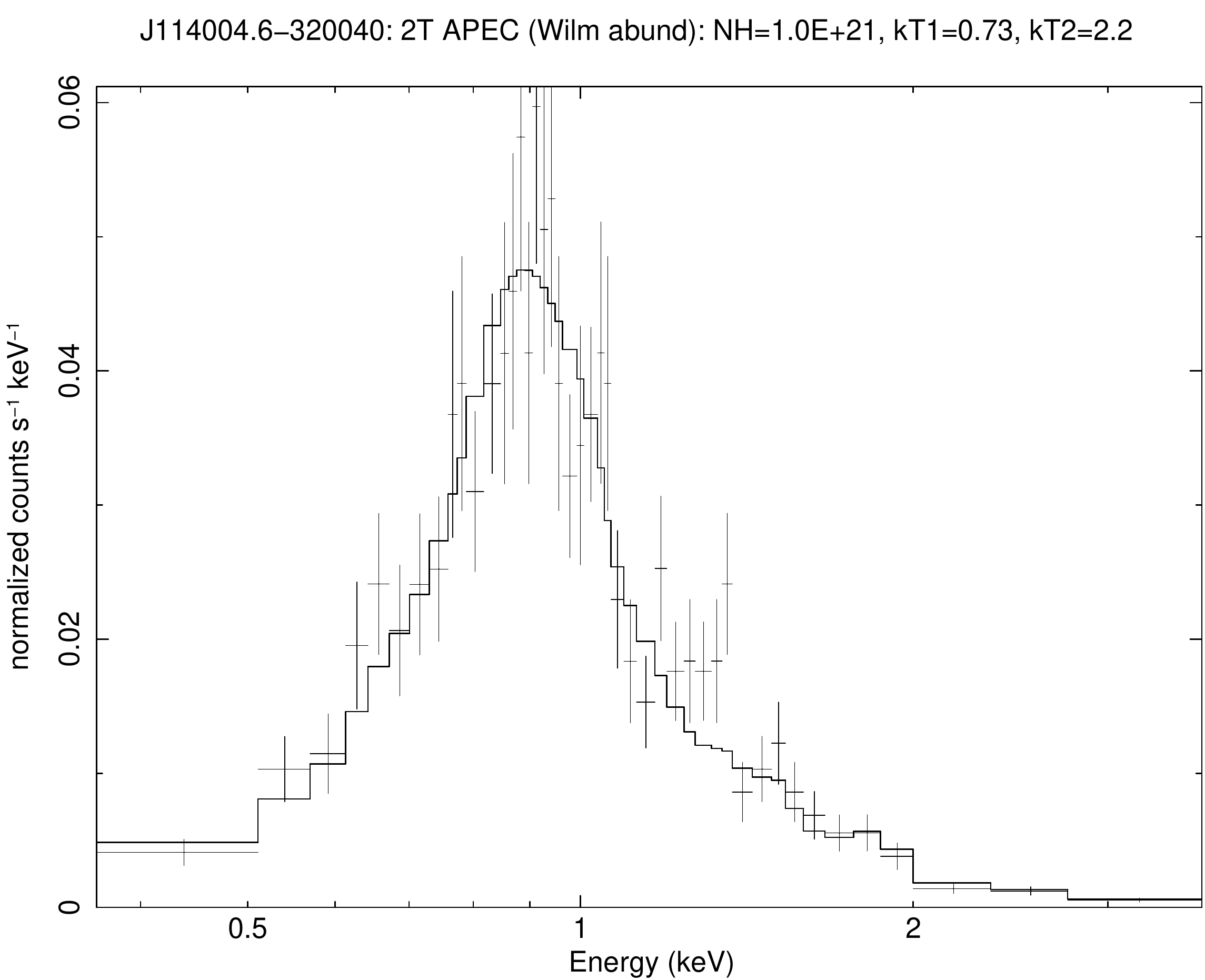}
\includegraphics[width=2.5in,angle=270]{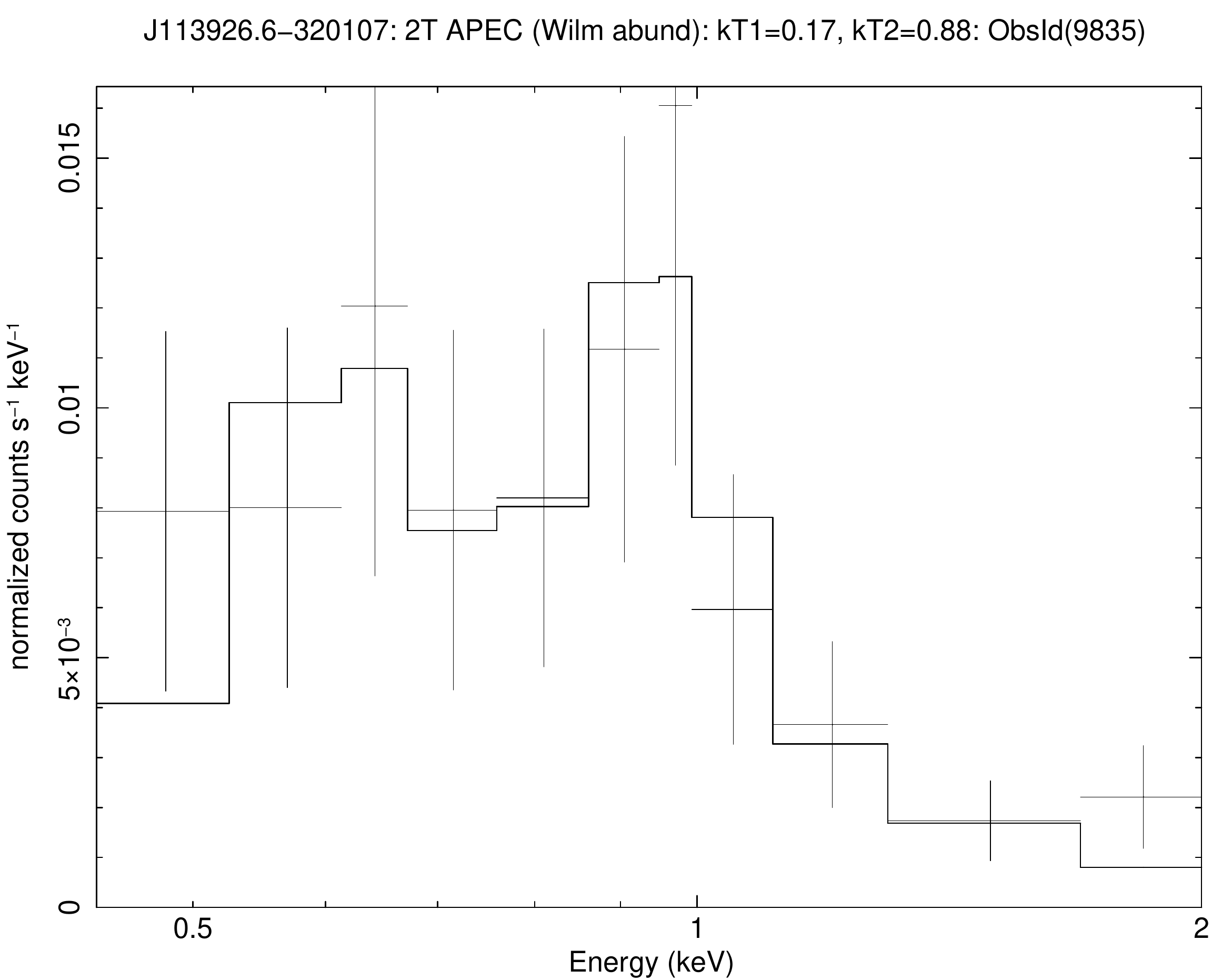}
\includegraphics[width=2.5in,angle=270]{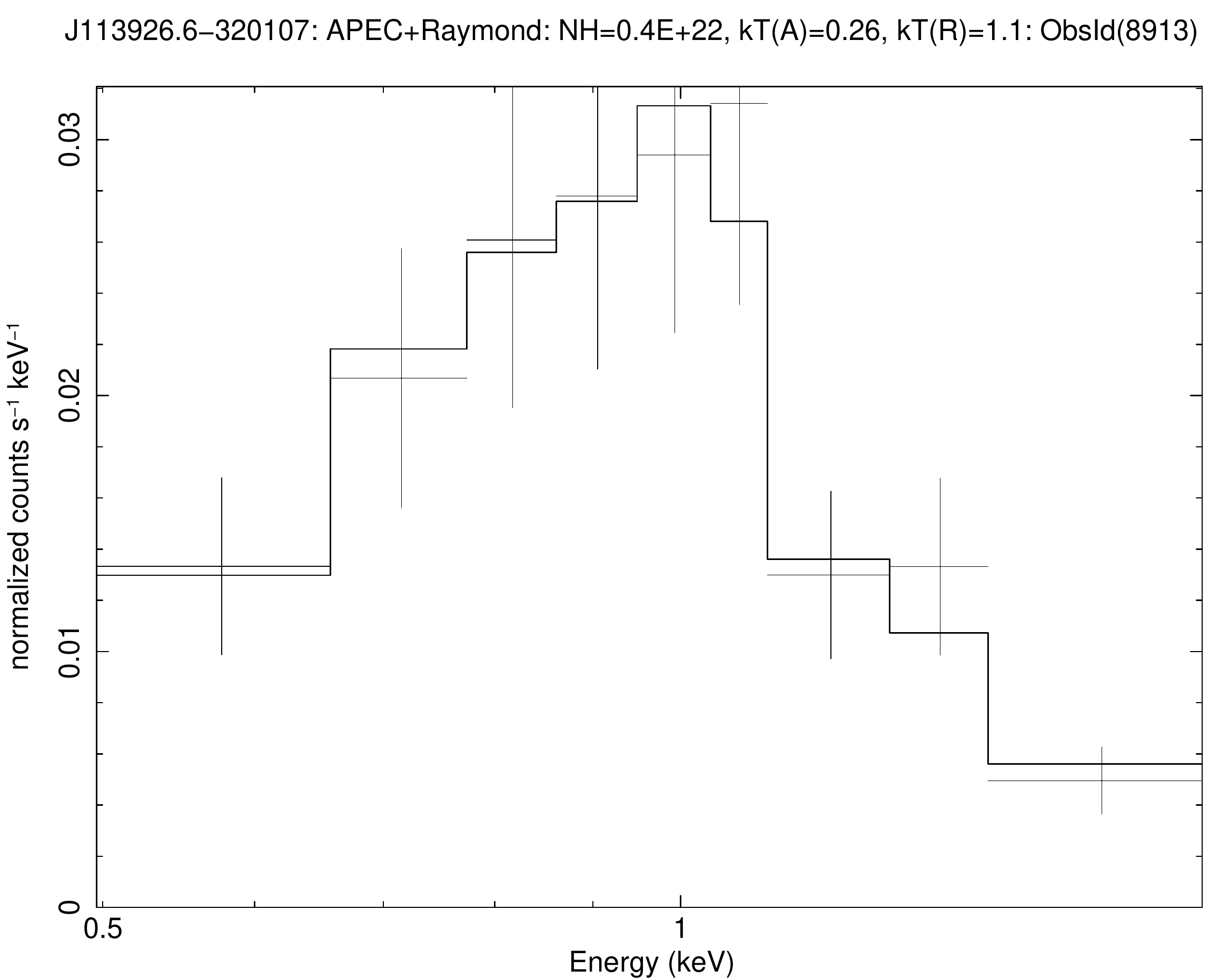}
\end{figure*}

This object was observed at the CTIO by the 1.5m spectrograph using grating 32/I, over a wavelength of 
$\sim6000$ to $9000$ \AA, with a resolution of $\sim3$ \AA$\phantom{0}$pixel$^{-1}$. Four observations were 
obtained on March 10, 2010, with an exposure time of 100 seconds each. The data was reduced with standard 
IRAF procedures and were combined to obtain the best signal to noise. 

We used the spectral standards of \citet{DanksDennefeld1994} which covered our entire wavelength range and 
was similar to our resolution at 4 \AA$\phantom{0}$pixel$^{-1}$. We used absorption features of Ca I at 6162 \AA, 
the blended line of Ba II, Fe I, and Ca I at 6497 \AA, and the Mg I line at 8807 \AA$\phantom{0}$to determine 
the spectral type of J114004.6-320040 \citep{DanksDennefeld1994,Allenetal1995,Torres-Dodgenetal1993}. 
The H$\alpha$ line at 6563 \AA, and the Ca II triplet at 8498, 8542, and 8662 \AA, all show filling, which is 
indicative of chromospheric activity, refer to Figure 6. The Ca II triplet is formed in the lower chromosphere 
and shares the upper level of the Ca II H \& K transitions, which are the most widely used optical indicators of 
chromospheric activity. The H$\alpha$ line is formed at the middle chromosphere and in less active stars only 
has a filled in absorption line \citep{Montesetal2004,Mallik1994,Mallik1997}.

Based on the absorption spectral features mentioned above in conjunction with the slope of the flux calibrated 
spectra continuum to that of the standard, we find this object to be of spectral type K0IV. The flux calibrated 
spectra is consistent with the standard but shows interstellar reddening (refer to Figure 6). With this spectral 
classification, we have $\log L_{\rm X}/L_{\rm bol}=-3.6$, using BC$_{V}=-0.4$ which is the average of the BC$_{V}$
for the main sequence K0V, which gives $\log L_{\rm X}/L_{\rm bol}=-3.5$, and the giant K0III, which gives 
$\log L_{\rm X}/L_{\rm bol}=-3.6$ \citep{Allens2000}.

\begin{figure*}
\caption{Spectra of J114004.6-320040 (thick black line) with comparison standard spectra (thin red line)
of \protect\citet{DanksDennefeld1994}. Top: Normalized spectra showing the best fit standard spectra of K0IV. 
Absorption features that were used to classify the spectral type and those that show filling are labeled.
Bottom: Flux calibrated spectra showing interstellar reddening of J114004.6-320040 compared to the standard. 
}
\includegraphics[width=5.in]{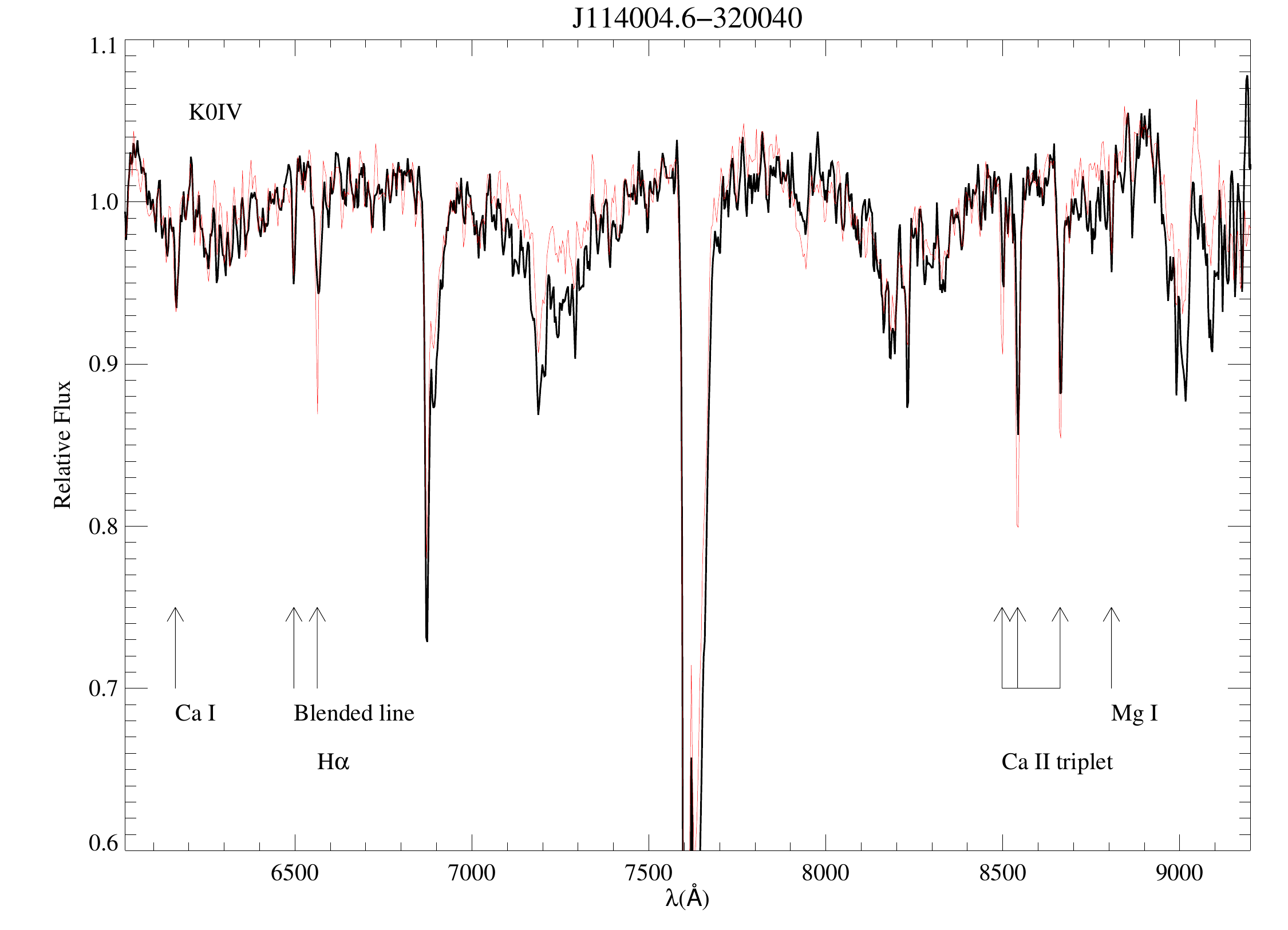}
\includegraphics[width=5.in]{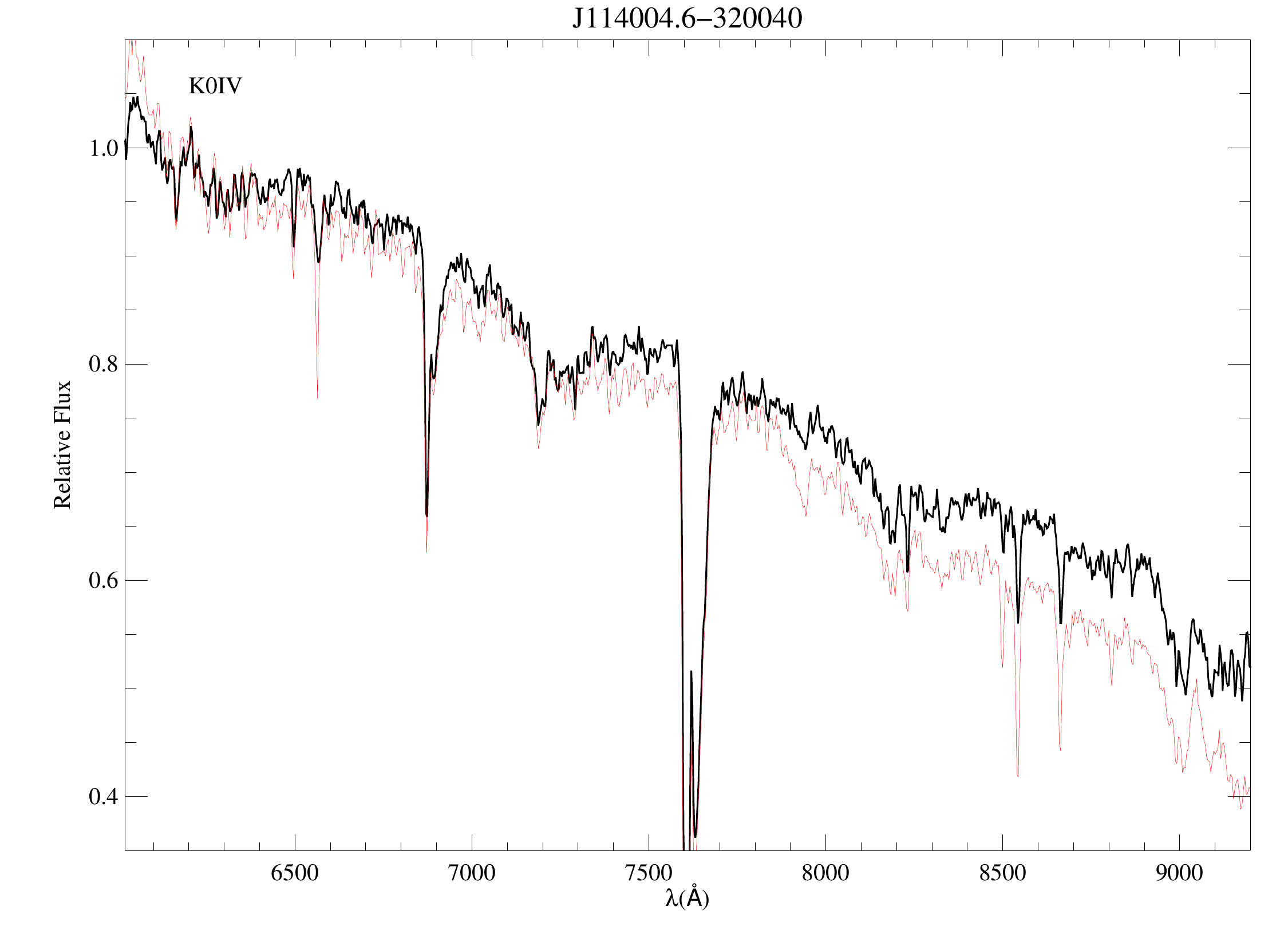}
\end{figure*}

Single subgiants are generally not known to be strong X-ray emitters, and overall show low activity 
\citep{Dorrenetal1995,Walteretal1980}. One of the nearest subgiants, $\beta$ Hyi, was found by ROSAT to have 
an X-ray luminosity of $L_{\rm X}=(0.9 - 3.0)$ x $10^{27}$ ergs s$^{-1}$ \citep{Dorrenetal1995}. As stars 
evolve their rotation rates decrease through magnetic braking which is accompanied by a parallel decrease in 
the various measures of stellar activity \citep{Dempseyetal1993}. However, RS Canum Venaticorum (RS CVn) binary 
systems are of the most luminous stellar coronal X-ray sources, having X-ray flux orders of magnitude more than 
slowly rotating single late-type stars, $L_{\rm X}\sim10^{29-31.5}$ ergs s$^{-1}$ \citep{Ostenetal2000,Dempseyetal1993}.
RS CVns usually consist of a G or K-type giant or subgiant with a late-type main sequence or subgiant companion, 
where the definition of these systems was first defined by \citet{Hall1976} \citep{Dempseyetal1993}. 
These are systems that have fast rotation because tidal forces have caused the rotation period to be 
synchronized with the orbital period, $\le30$ days, where this rapid rotation enhances stellar 
activity \citep{Ostenetal2000,Dempseyetal1993}. We classify J114004.6-320040 as a RS CVn, K0IV.
We saw no evidence of binarity based on our low-dispersion spectra. Follow-up, higher-dispersion 
(preferably echelle) spectral monitoring is needed to confirm this object as a bona fide RS CVn.

\subsection{J113926.6-320107}
We analyze the second brightest object in the FOV of our {\it Chandra} observations, with almost 200 counts.
This object was found in the 2MASS catalog and was detected by GALEX in the NUV and the FUV. No object was 
found in the SIMBAD database within $2^{\prime \prime}$ of the R.A. and Decl.

The light-curve for J113926.6-320107 is shown in Figure 4 (bottom) with clear variability. ObsId 9835 has 
a KS probability of 0.4, the object is considered quiescent during this observation. ObsId 8913 has a KS 
probability of 8 x $10^{-7}$, the light-curve with great confidence is variable. Qualitatively, ObsId 8913 
clearly shows the decay phase of a flare whose rise has occurred at some time between ObsId 9835 and ObsId 8913.

We analyze the spectra of J113926.6-320107 separately for ObsId 9835 and ObsId 8913 since the former is of 
quiescent emission and the latter is during the decay phase of a flare. We fit the spectra of 
ObsId 9835 (Figure 5, top right) with a two temperature APEC (Wilm abund) model. We find a good fit with
$kT_{1}=0.17\pm0.05$ keV, $kT_{2}=0.88\pm0.23$ keV, and a fixed column density from ObsId 8913 of 
$N_{\rm H}=0.4\pm0.4$ x $10^{22}$ cm$^{-2}$. We find an unabsorbed flux of 
$f_{\rm xu}=1.3$ x $10^{-13}$ ergs cm$^{-2}$ s$^{-1}$ and an absorbed flux of 
$f_{\rm xa}=3.3^{+0.3}_{-1.8}$ x $10^{-14}$ ergs cm$^{-2}$ s$^{-1}$. We fit the spectra of 
ObsId 8913 (Figure 5, bottom) with a two temperature APEC+Raymond (Wilm abund) model. We find a good fit with
$kT_{\rm APEC}=0.26\pm0.10$ keV, $kT_{\rm Raymond}=1.1\pm0.1$ keV, and a column density of 
$N_{\rm H}=0.4\pm0.4$ x $10^{22}$ cm$^{-2}$. We find an unabsorbed flux of 
$f_{\rm xu}=2.4$ x $10^{-13}$ ergs cm$^{-2}$ s$^{-1}$ and an absorbed flux of 
$f_{\rm xa}=9.7^{+0.3}_{-4.9}$ x $10^{-14}$ ergs cm$^{-2}$ s$^{-1}$.

\bibliographystyle{apj}
\bibliography{/Users/Phil/research/bibliography}

\label{lastpage}

\end{document}